\documentclass[aps,prd,twocolumn,amsmath,amssymb,groupedaddress,tightenlines,showpacs,preprintnumbers]{revtex4-1}

\usepackage{cancel}
\usepackage[pdftex]{graphicx}
\usepackage[pdftex, pdfstartview={FitH}, pdfnewwindow=true, colorlinks=true, citecolor=blue, filecolor=blue, linkcolor=blue, urlcolor=blue, pdfpagemode=UseNone, bookmarks=false]{hyperref}

\def\PRDheight{5.36cm}
\newcommand{\al}{\ensuremath{\alpha} }
\newcommand{\be}{\ensuremath{\beta} }
\newcommand{\ga}{\ensuremath{\gamma} }
\newcommand{\La}{\ensuremath{\Lambda} }
\newcommand{\la}{\ensuremath{\lambda} }

\newcommand{\chibar}{\ensuremath{\overline\chi} }
\newcommand{\psibar}{\ensuremath{\overline\psi} }
\newcommand{\vev}[1]{\ensuremath{\left\langle #1 \right\rangle} }
\newcommand{\pbp}{\ensuremath{\vev{\psibar\psi}} }
\newcommand{\X}{\ensuremath{\!\times\!} }

\renewcommand{\tilde}[1]{\ensuremath{\widetilde #1 } }
\newcommand{\secref}[1]{Section~\ref{#1}}
\newcommand{\refcite}[1]{Ref.~\cite{#1}}
\newcommand{\eqn}[1]{Eqn.~\ref{#1}}
\newcommand{\fig}[1]{Fig.~\ref{#1}}
\newcommand{\Sb}{\ensuremath{\cancel{S^4}} }

\def\ReTr{{\rm Re \, Tr}\,}

\begin{document}
\title{Novel phase in SU(3) lattice gauge theory with 12 light fermions}

\author{Anqi~Cheng}
\affiliation{Department of Physics, University of Colorado, Boulder, CO 80309, USA}
\author{Anna~Hasenfratz}
\affiliation{Department of Physics, University of Colorado, Boulder, CO 80309, USA}
\author{David~Schaich}
\affiliation{Department of Physics, University of Colorado, Boulder, CO 80309, USA}

\date{29 March 2012}

\begin{abstract} 
  We study the phase structure of SU(3) lattice gauge theory with $N_f = 12$ staggered fermions in the fundamental representation, for both zero and finite temperature at strong gauge couplings.
  For small fermion masses we find two transitions at finite temperature that converge to two well-separated bulk phase transitions.
  The phase between the two transitions appears to be a novel phase.
  We identify order parameters showing that the single-site shift symmetry of staggered fermions is spontaneously broken in this phase.
  We investigate the eigenvalue spectrum of the Dirac operator, the static potential and the meson spectrum, which collectively establish that this novel phase is confining but chirally symmetric.
  The phase is bordered by first-order phase transitions, and since we find the same phase structure with $N_f = 8$ fermions, we argue that the novel phase is most likely a strong-coupling lattice artifact, the existence of which does not imply IR conformality.
\end{abstract}
\preprint{Colo-HEP 572}
\maketitle

\section{Introduction} 
Strongly coupled gauge--fermion systems, beyond their intrinsic theoretical interest, play an essential role in many theories of physics beyond the standard model~\cite{Rychkov:2011br}.
Lattice gauge theory is at present the most reliable method to study strongly-interacting models in a systematic, controlled way.
Most lattice studies have focused on determining whether given models exhibit confinement and chiral symmetry breaking, or if they develop an infrared fixed point (IRFP) resulting in IR conformality~\cite{DelDebbio:2011rc}.
SU(3) gauge theory with $N_f = 12$ fundamental flavors has emerged in lattice studies as one of the most controversial models~\cite{Appelquist:2009ty, Deuzeman:2009mh, Jin:2009mc, Hasenfratz:2010fi, Deuzeman:2010fn, Fodor:2011tu, Hasenfratz:2011xn, Appelquist:2011dp, DeGrand:2011cu, Aoyama:2011ry, Ogawa:2011ki, Deuzeman:2011pa, Hasenfratz:2011da, Deuzeman:2012pv, Aoki:2012kr, Jin:2012dw}.

A recent large-scale study of the $N_f = 12$ system concluded that the data favored a confining, chirally broken scenario~\cite{Fodor:2011tu}, though other groups interpreted these data as consistent with IR conformality~\cite{Appelquist:2011dp, DeGrand:2011cu}.
One of us recently investigated the same system with an entirely different approach, the Monte Carlo renormalization group (MCRG) two-lattice matching method~\cite{Hasenfratz:2010fi, Hasenfratz:2011xn, Hasenfratz:2011da}, finding an IRFP consistent with IR-conformal dynamics.
An obvious difference between \refcite{Fodor:2011tu} and \refcite{Hasenfratz:2011xn} is that the MCRG analysis found the IRFP at a considerably weaker bare coupling than that considered in \refcite{Fodor:2011tu}.

In this work we present a study of the $N_f = 12$ system at stronger couplings, reporting results for the phase diagram in the gauge coupling--fermion mass parameter space, at both zero and finite temperature.
We find two transitions at finite temperature that converge to two well-separated bulk phase transitions, consistent with what Refs.~\cite{Deuzeman:2010fn, Schroeder:2011LAT} observed using different staggered lattice actions.
The general consistency of results obtained with very different actions indicates that we are observing a robust feature of lattice gauge theories with many staggered fermions.
Shortly after we completed this work, \refcite{Deuzeman:2011pa} interpreted the second transition as a partial restoration of axial U(1)$_A$ symmetry, which is not consistent with our data discussed below.

Between the two bulk transitions we identify a novel phase, and will devote most of this paper to understanding it. We will show that this novel phase breaks the single-site shift symmetry of the staggered action, a property that to our knowledge has never been observed before.
It is important to establish whether or not this phase is a lattice artifact, because this will affect the conclusions we can draw from our observation of finite-temperature transitions converging to bulk transitions.
As we will discuss in \secref{sec:phase}, such behavior is characteristic of IR-conformal systems, and in principle provides a signal that distinguishes confining and conformal systems~\cite{Deuzeman:2009mh}.
However, if the phase is a lattice artifact such as the ``Aoki-like phase for staggered fermions'' discussed by \refcite{Aubin:2004dm}, then it would be bounded by bulk transitions regardless of whether the 12-flavor system were confining or conformal in the continuum.
Moreover, some finite-temperature transitions could converge to these bulk transitions for both confining and conformal systems.
As a consequence, our results would not necessarily imply (although they would be consistent with) IR-conformal continuum dynamics.

The outline and main results of our paper are as follows.
After summarizing our lattice action in \secref{sec:action}, we present our results for the phase structure at light fermion mass in \secref{sec:phase}.
Using two different order parameters, in \secref{sec:order} we will clearly show that the single-site shift symmetry (``$S^4$'') of the staggered action is spontaneously broken in the novel (``\Sb'') phase.
In \secref{sec:eigen} we study the low-lying eigenvalues of the Dirac operator in both the \Sb phase as well as the more familiar weak-coupling phase.
The volume scaling of the low-lying eigenmodes in the \Sb phase indicates the presence of a ``soft edge'' $\la_0 > 0$ in the density distribution, $\rho(\lambda) \propto (\lambda - \lambda_0)^{\alpha}$.
A soft edge implies that not only the chiral condensate $\pbp$, but also higher-order condensates that could break chiral symmetry, are vanishing in the chiral limit~\cite{Damgaard:2000cx, Chandrasekharan:1998yx}.
In the weak-coupling phase, we do not observe a soft edge in the eigenvalue density distribution.
We also obtain a preliminary prediction for the mass anomalous dimension $\gamma_m = 0.61(5)$, where the error is purely statistical.

We present more evidence that the \Sb phase is chirally symmetric in \secref{sec:pot_spec}, through the light meson spectrum.
In the \Sb phase we observe parity doubling between scalar and pseudoscalar states as well as between vector and axial-vector states.
Our meson spectrum results in the \Sb phase are also independent of the volume, in contrast to those in the weak-coupling phase.
We also investigate the static potential, which shows that the \Sb phase is confining.
The static potential predicts a non-vanishing string tension and a small Sommer parameter $r_0 \approx 2.7$.
Minimal finite volume effects both in the meson spectrum and static potential are consistent with this $r_0$, and show that the lattice correlation length is small in \Sb phase.

To explore whether the \Sb phase could exist without implying IR conformality in the continuum, we have performed preliminary investigations of the phase structure with $N_f = 8$ fundamental flavors, a system generally believed to show spontaneous chiral symmetry breaking.
While we have not yet completed our exploration of the $N_f = 8$ phase diagram, our initial results in \secref{sec:8f} show an \Sb phase with the same properties as we observe with 12 flavors.
We do not observe this phase with 4 flavors, and our data with 16 flavors are currently too preliminary to draw a definite conclusion.
Our investigations with 8 and 16 flavors are ongoing, and we will report our full results in future works~\cite{Cheng:2012TBD}.

We conclude in \secref{sec:conclusions}.
Although we find the \Sb phase to be confining yet chirally symmetric, it is bordered by first-order phase transitions that prevent us from reaching the continuum (infinite cut-off) limit.
It is possible that the \Sb phase is a lattice artifact of staggered fermions, especially since it exists with 8 flavors as well.  This prevents us from interpreting our observation of finite-temperature transitions converging to bulk transitions as evidence that the 12-flavor system is IR-conformal.
While \refcite{Aubin:2004dm} has discussed a possible Aoki-like lattice artifact phase for staggered fermions, it is not clear if the breaking of the single-site shift symmetry we observe in the \Sb phase can be described by the staggered chiral Lagrangian.
This question deserves further study, but it is beyond the scope of the present paper.

\section{\label{sec:action} The lattice action} 
Lattice calculations are affected by discretization errors, and much effort has been devoted to improving lattice actions to reduce these effects.
Strongly-coupled systems are particularly sensitive to these lattice artifacts, which can contaminate or destroy the scaling of the desired continuum limit, even to the point of generating spurious ultraviolet fixed points (UVFPs).
Care must be taken that lattice simulations are in the basin of attraction of the perturbative fixed point, or its associated IRFP if it exists.

In \refcite{Hasenfratz:2011xn} we advocated the use of a gauge action with a negative adjoint plaquette term, to avoid a well-known spurious UVFP caused by lattice artifacts.
In the present work we follow this suggestion and use a gauge action that includes both fundamental and adjoint plaquette terms, with coefficients related by $\be_A = -0.25\be_F$.
With this constant ratio, the perturbative relation to the bare gauge coupling is
\begin{equation}
  6 / g^2 = \be_F(1 + 2\be_A / \be_F) = \be_F / 2.
\end{equation}

Staggered fermions are also affected by taste breaking, i.e., the four fermion tastes described by each (unrooted) staggered field are degenerate only in the continuum limit.
Smearing the gauge connections reduces this problem, and following Refs.~\cite{Hasenfratz:2009ea, Hasenfratz:2010fi, Hasenfratz:2011xn}, we use nHYP-smeared staggered fermions.
The nHYP smearing significantly improves the taste symmetry of staggered fermions~\cite{Hasenfratz:2001hp, Hasenfratz:2007rf}, but the U(3) projection in the nHYP construction can break down at strong coupling, due to the generation of near-zero eigenvalues in the staple sum.
We address this difficulty by adjusting the three HYP smearing parameters to
\begin{equation*}
  \al_1 = 0.5, \quad \al_2 = 0.5, \quad \al_3 = 0.4
\end{equation*}
from the original $(0.75, 0.6, 0.3)$ values.
This choice eliminates the numerical problems while it only slightly increases taste splitting.

In our calculations we use the hybrid Monte Carlo (HMC) algorithm.
Our code is based in part on the MILC Collaboration's public lattice gauge theory software~\footnote{http://www.physics.utah.edu/$\sim$detar/milc/}.
We have modified this software to implement nHYP smearing, to add the adjoint plaquette term to the gauge action, and to exploit both even and odd sublattices to simulate eight flavors.
During the course of our work, we also implemented a second-order Omelyan integrator~\cite{Takaishi:2005tz} accelerated by an additional heavy pseudofermion field~\cite{Hasenbusch:2002ai} and multiple time scales~\cite{Urbach:2005ji}.
Our HMC trajectory length is typically one molecular dynamics time unit (MDTU), but in some cases can be as small as 0.5 MDTU or as large as 2.0 MDTU.
For most of the ensembles discussed in this work, we accumulated 1000--2000 MDTU, and measured the eigenvalues and meson spectrum on every tenth trajectory.
Around the phase transitions we accumulated more than 10,000 MDTU for some ensembles.

\section{\label{sec:phase}The phase structure} 
\begin{figure}
  \centering
  \includegraphics[height=\PRDheight]{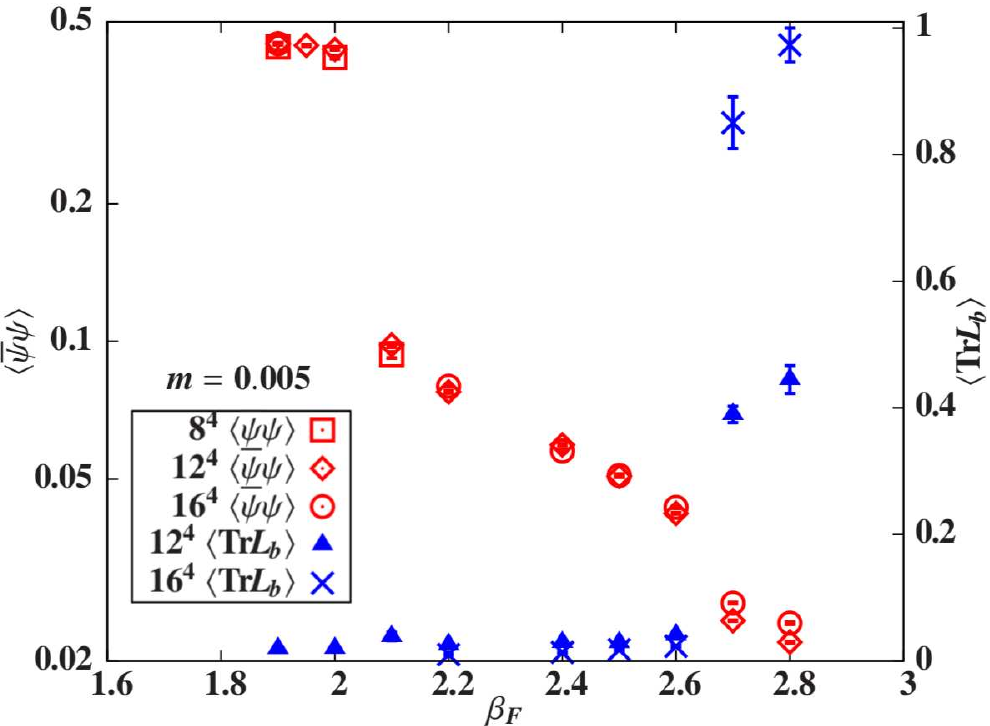}
  \caption{\label{fig:pbp_Lb} The chiral condensate \pbp (on a log scale) and the blocked Polyakov loop $\vev{\mbox{Tr}L_b}$ indicate two well-separated transitions at $m=0.005$.}
\end{figure}

\begin{figure*}[ht]
  \centering
  \includegraphics[width=0.32\linewidth]{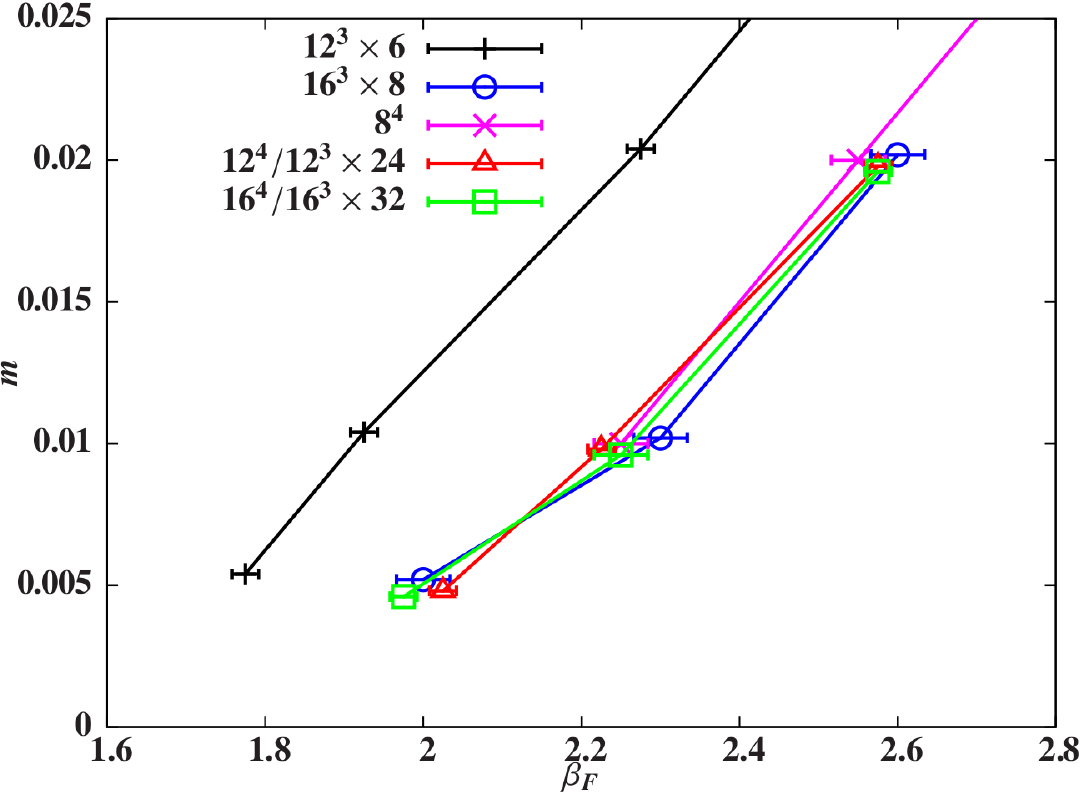}\hfill
  \includegraphics[width=0.32\linewidth]{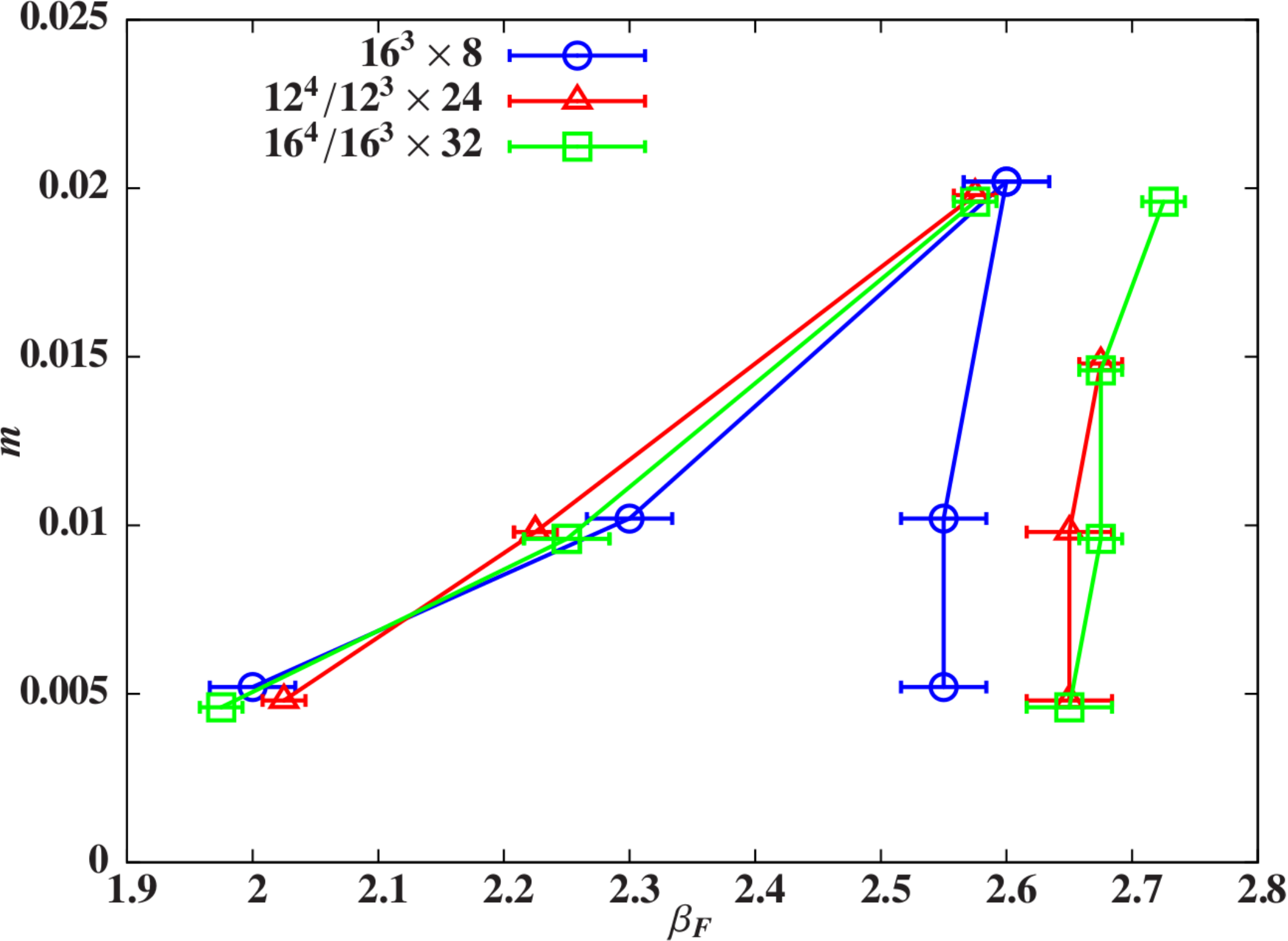}\hfill
  \includegraphics[width=0.32\linewidth]{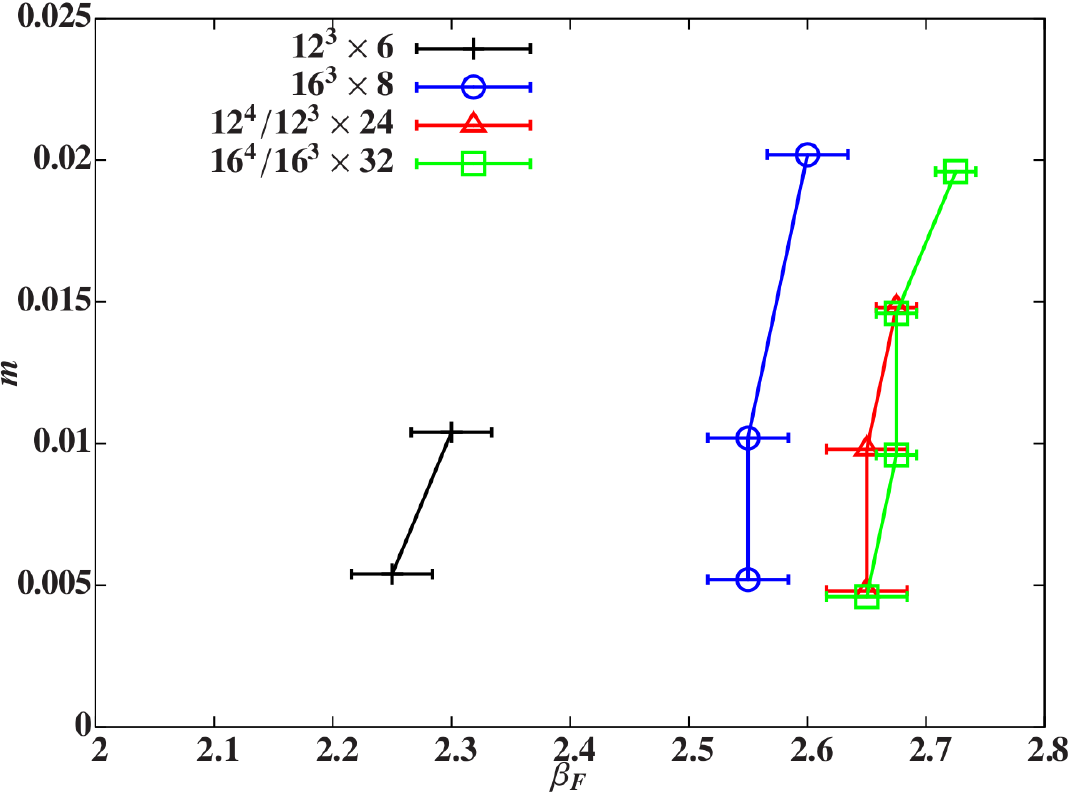}
  \caption{\label{fig:transitions} Transitions in the $\be_F$--$m$ plane at several temperatures and volumes.  The left panel shows the stronger-coupling transitions signaled by $\pbp$, the right panel shows the weaker-coupling transitions signaled by $\vev{\mbox{Tr}L_b}$, and the middle panel shows how the two bulk transitions merge as $m$ increases.  Small vertical offsets distinguish the different volumes, and lines connect the points to guide the eye.  The transitions are nearly identical on zero-temperature volumes, and the finite-temperature transitions appear to converge to these bulk transitions.}
\end{figure*}

In the $m = 0$ chiral limit, confining and chirally broken systems with $N_f \ge 3$ flavors of fundamental fermions are expected to exhibit a first-order finite-temperature phase transition at which they become chirally symmetric and deconfined.
A finite-temperature lattice system with fixed $N_t \ll L$ will undergo a phase transition at a critical coupling $\be_F^{(c)}$.
In the weak-coupling scaling region the renormalization group equation predicts the dependence of $\be_F^{(c)}$ on $N_t$.
In order for the theory to be confining and chirally broken at zero temperature, $\be_F^{(c)} \to \infty$ as $N_t \to \infty$.

Zero-temperature systems with $N_t \ge L$ deconfine and become chirally symmetric when $L$ is so small that the physics is volume-squeezed.
This is a finite-volume effect and not a real phase transition, though it could be accompanied by a discontinuity.
In few-flavor QCD-like systems no discontinuity is observed at zero temperature.

Much less is known about the finite-temperature behavior of IR-conformal systems.
At strong enough coupling, lattice artifacts create a confining, chirally broken phase on the lattice~\footnote{There is no rigorous proof of this statement but all investigated models, including the 12-flavor SU(3) system, show this behavior.}.
This strong-coupling phase must be separated from the weak-coupling conformal phase by a bulk (non-thermal) phase transition in the chiral limit.
The bulk transition has to be a real infinite-volume transition, with the chiral condensate \pbp serving as an order parameter in the chiral limit.
While remnants of the finite-temperature phase transition can coexist with the bulk transition, the finite-temperature transitions must occur at stronger couplings than the bulk transition, and converge to the bulk transition as $N_t \to \infty$.
This, in principle, gives a signal that distinguishes confining and conformal systems.

Refs.~\cite{Deuzeman:2009mh, Deuzeman:2010fn, Deuzeman:2011pa} investigated the 12-flavor SU(3) system and found indication for a bulk transition.
Refs.~\cite{Deuzeman:2010fn, Deuzeman:2011pa, Schroeder:2011LAT} also discussed a second discontinuity in $\pbp$.
Our investigations confirm the existence of both bulk phase transitions, as illustrated in \fig{fig:pbp_Lb}.
In the chiral condensate we observe a clear discontinuity around $\be_F \approx 2.0$ for $m = 0.005$ on zero-temperature $8^4$, $12^4$ and $16^4$ volumes.
\pbp has another, much smaller, discontinuity around $\be_F \approx 2.65$, where the Polyakov loop, an observable related to confinement, shows a much stronger signal.
Because the usual Polyakov loop becomes small and noisy as $N_t$ increases, we consider an improved observable by measuring the Polyakov loop on renormalization group blocked lattices.
This blocked Polyakov loop $\vev{\mbox{Tr}L_b}$ has the same $Z_3$ symmetry as the standard one, and can also be thought of as an extended observable on the original, unblocked lattices.

The chiral condensate \pbp has very little volume dependence across the phase transitions, consistent with bulk transitions.
The apparent volume dependence of the blocked Polyakov loop is due to the different number of blocking steps performed: the $16^4$ lattices are blocked three times with scale factor $s = 2$, while the $12^4$ lattices can be blocked only twice.

\fig{fig:transitions} shows how the locations of the two phase transitions depend on the volume, temperature and fermion mass.
In both cases, the transitions at finite temperature converge to zero-temperature bulk transitions where different observables show the same discontinuity on all volumes, up to small finite volume effects.
Just like Refs.~\cite{Deuzeman:2009mh, Deuzeman:2010fn, Deuzeman:2011pa}, we observe the stronger-coupling transitions to converge on smaller volumes than those that are needed for the weaker-coupling transitions to converge.
We encountered long metastability between runs from hot and cold initial states at both transitions, especially on larger volumes.
These transitions are strongly first-order, and molecular dynamics evolution is not very effective flipping the system between phases.
Mixed initial configurations helped to resolve the transition around $\be_F \approx 2.65$ more accurately, but they were less reliable at the transition around $\be_F \approx 2$.

\section{\label{sec:order}Single-site shift symmetry breaking} 
\begin{figure*}
  \centering
  \includegraphics[width=0.45\linewidth]{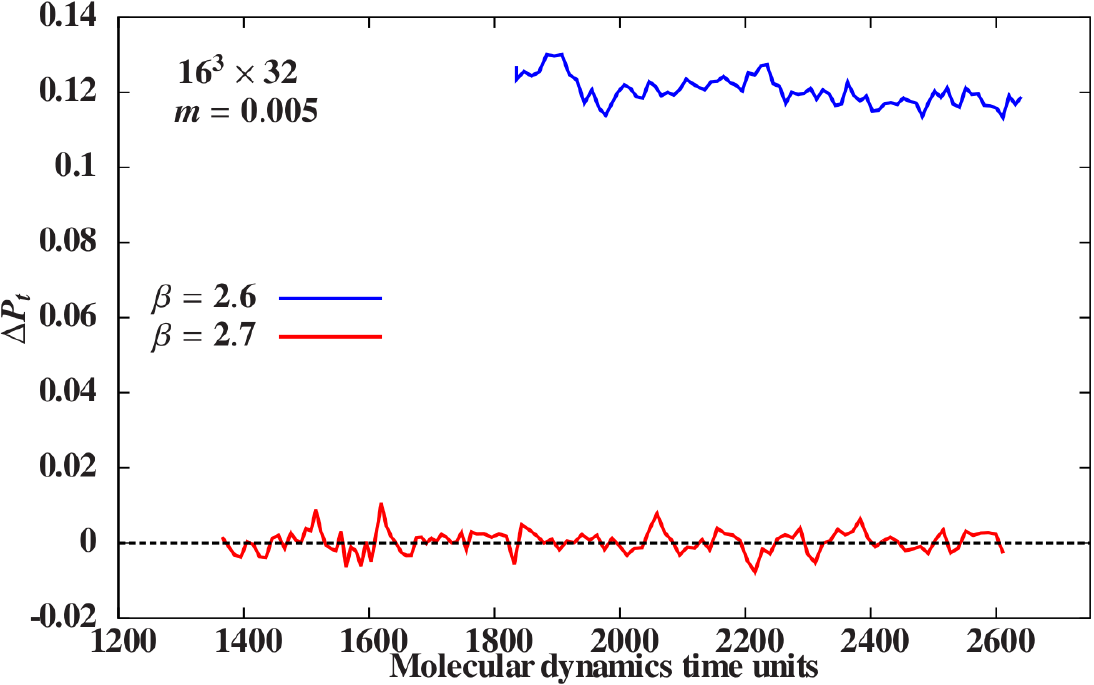}\hfill
  \includegraphics[width=0.45\linewidth]{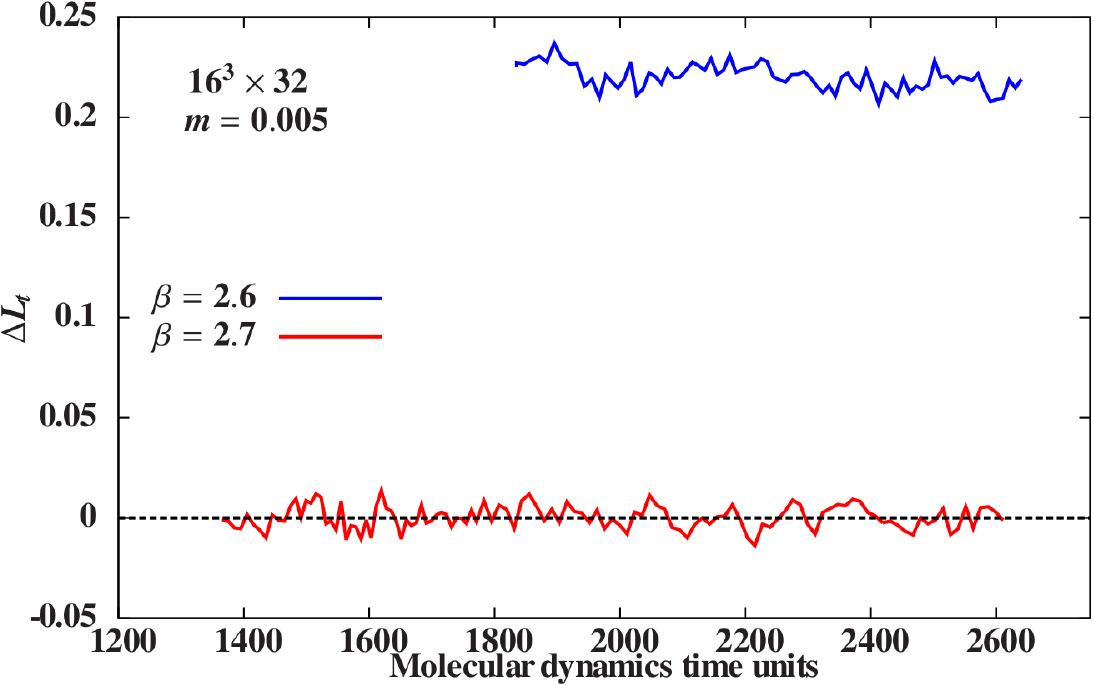}
  \caption{\label{fig:order_param} The plaquette difference $\Delta P_t$ (\eqn{eq:plaq_diff}, left) and the link difference $\Delta L_t$ (\eqn{eq:link_diff}, right) measured on $16\X32$ volumes in both the \Sb phase ($\beta_F=2.6$, $m=0.005$) and the weak-coupling phase ($\beta_F=2.7$, $m=0.005$), as functions of the molecular dynamics time.}
\end{figure*}

We identified two phase transitions in \fig{fig:pbp_Lb} from the discontinuities in the chiral condensate and (blocked) Polyakov loop.
This was possible as both phase transitions are first-order, and almost all observables show a discontinuity.
However, neither $\pbp$ nor the Polyakov loop is a {\em bona fide} order parameter of the intermediate phase located between the two transitions.
While the Polyakov loop is a good indicator of confinement, it is only an order parameter in the pure gauge theory, and does not distinguish between the intermediate and strong-coupling phases.
The chiral condensate is an order parameter in the chiral limit only, and in that limit it likely vanishes in both the intermediate and weak-coupling phases.

There is no {\em a priori} guarantee that the intermediate phase is separated from the strong- and weak-coupling phases by true phase transitions.
However, while investigating the phase diagram, we discovered that the single-site shift symmetry (``$S^4$'') of the staggered fermion action is spontaneously broken in the intermediate (``\Sb'') phase.
This ensures the existence of an order parameter characterizing the \Sb phase, and full separation of the phases.

The single-site shift symmetry of the staggered action takes the form~\cite{Golterman:1984cy}
\begin{align}
  \chi(n) & \to \xi_{\mu}(n) \chi(n + \mu), & \chibar(n) & \to \xi_{\mu}(n) \chibar(n + \mu), \nonumber \\
  U_{\mu}(n) & \to U_{\mu}(n + \mu), & & \label{eq:shift_sym}
\end{align}
where
\begin{equation*}
  \xi_\mu(n) = (-1)^{\sum_{\nu > \mu} n_\nu}.
\end{equation*}
This symmetry ensures that the chiral condensate \pbp measured on even lattice sites is identical to that measured on odd sites, and the underlying gauge configurations exhibit the usual discrete translational symmetry.
To our knowledge the breaking of this symmetry has never been observed before.

Order parameters that are sensitive to this symmetry include the expectation value of the difference between neighboring plaquettes and that between neighboring links,
\begin{align}
  \Delta P_{\mu} = & \vev{\ReTr \square_n - \ReTr \square_{n + \mu}}_{n_\mu {\rm even}} \label{eq:plaq_diff} \\
  \Delta L_{\mu} = & \langle\alpha_{\mu}(n) \chibar(n) U_{\mu}(n) \chi(n + \mu) \label{eq:link_diff} \\
                   & - \alpha_{\mu}(n + \mu)\chibar(n + \mu) U_{\mu}(n+\mu) \chi(n + 2\mu)\rangle_{n_\mu {\rm even}} \nonumber
\end{align}
where $\ReTr \square_n$ is the real trace of the plaquette originating at site $n$,
\begin{equation*}
 \alpha_\mu(n) = (-1) ^{\sum_{\nu < \mu} n_\nu}
\end{equation*}
is the usual staggered phase factor, and the expectation value $\vev{\cdots}_{n_{\mu} {\rm even}}$ is taken only over sites whose $\mu$ component is even.
In the intermediate phase these operators develop non-zero expectation values in one or more directions $\mu$.
Occasionally the direction of the symmetry breaking changes, rotating in space.

\fig{fig:order_param} shows both order parameters in the intermediate ($\beta_F=2.6$, $m=0.005$) and the weak-coupling ($\beta_F=2.7$, $m=0.005$) phases, as functions of the molecular dynamics time on $16^3\X32$ volumes.
At $\beta_F=2.6$ $S^4$ is broken in the temporal direction and both order parameters remain small for $\mu\ne t$.
At $\beta_F=2.7$ the order parameters do not develop a non-zero expectation value in any direction.
The order parameters also vanish in the strong-coupling phase, though we do not include that data in \fig{fig:order_param}.
The single-site shift symmetry is broken only in the intermediate phase.
The order parameters, when non-vanishing, have only small dependence on the volume.

It is important to note that the single-site shift symmetry is an exact symmetry of the action even at finite fermion mass.
It is broken only spontaneously.
Both $\Delta P_{\mu}$ of \eqn{eq:plaq_diff} and $\Delta L_{\mu}$ of \eqn{eq:link_diff} are nonzero when the symmetry is broken and vanish when it is preserved.
The \Sb phase must be separated by true phase transitions from the strong- and weak-coupling phases where both order parameters vanish.

\section{\label{sec:eigen}Eigenvalue spectrum} 
\begin{figure*}[ht]
  \centering
  \includegraphics[width=0.45\linewidth]{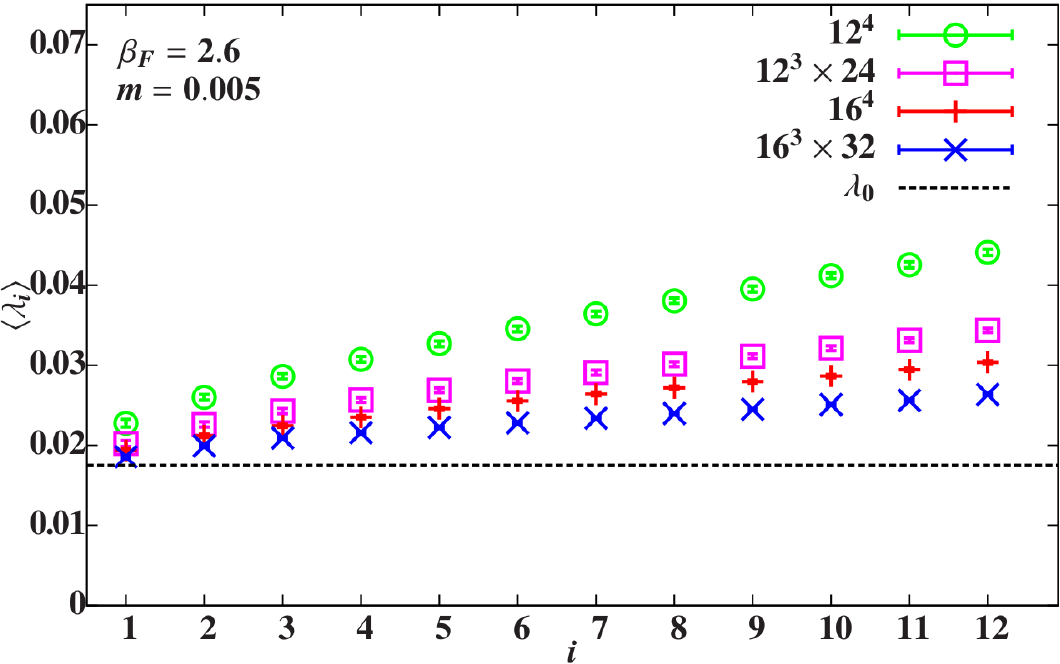}\hfill
  \includegraphics[width=0.45\linewidth]{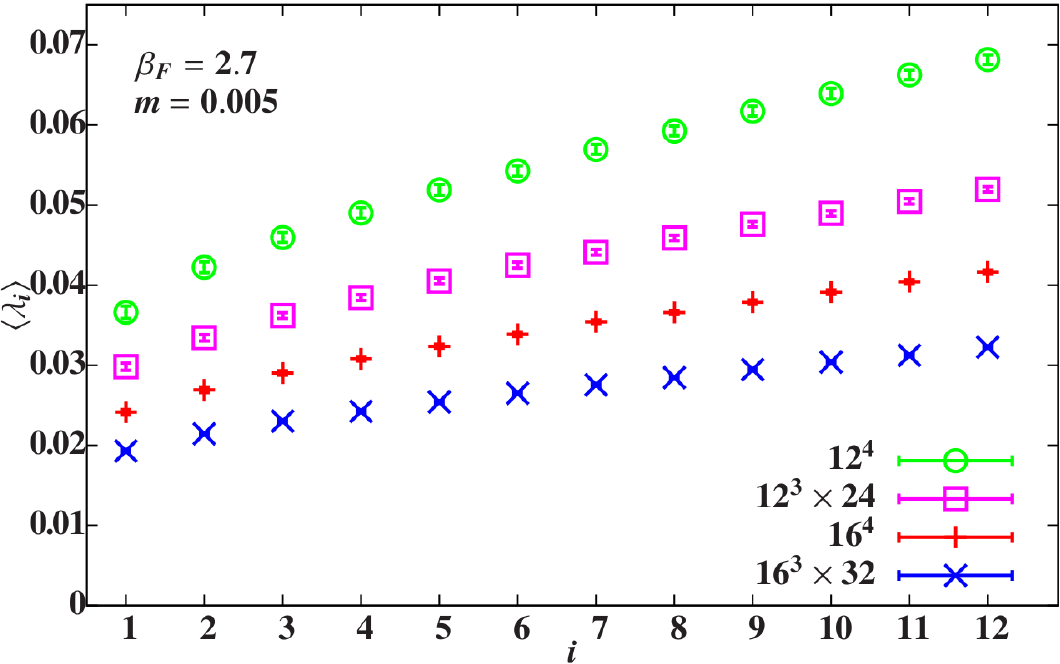}
  \caption{\label{fig:eigenvalues} The low-lying eigenvalues $\vev{\la_i}$ for $m = 0.005$ on the four volumes $12^4$, $12^3\X24$, $16^4$ and $16^3\X32$, in the \Sb phase ($\be_F = 2.6$, left) and in the weak-coupling phase ($\be_F = 2.7$, right).  The dashed line in the left panel shows the soft edge $\lambda_0 = 0.0175(5)$ found in the fit plotted in the left panel of \protect{\fig{fig:eig_scaling}}.}
\end{figure*}

\begin{figure*}
  \centering
  \includegraphics[width=0.45\linewidth]{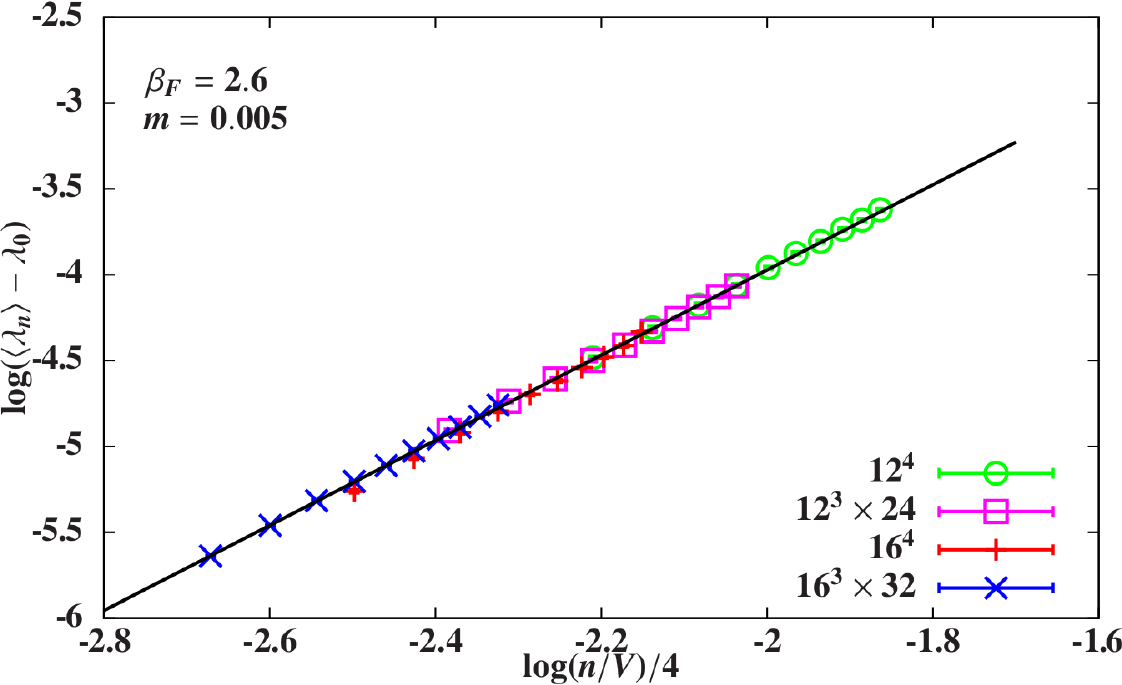}\hfill
  \includegraphics[width=0.45\linewidth]{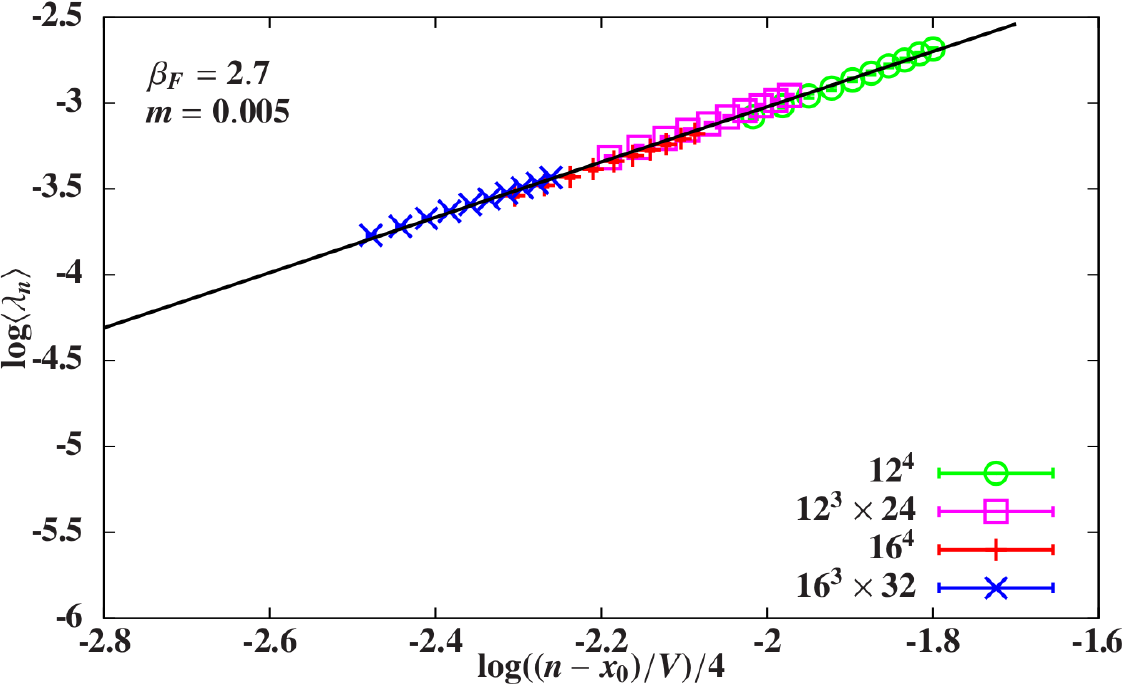}
  \caption{\label{fig:eig_scaling} Scaling of the low-lying eigenvalues for $m = 0.005$ on the four volumes $12^4$, $12^3\X24$, $16^4$ and $16^3\X32$.  In the \Sb phase ($\be_F = 2.6$, left) the soft edge $\lambda_0 = 0.0175(5)$, and the slope $y_m = 2.50(10)$.  In the weak-coupling phase ($\be_F = 2.7$, right), $\la_0 = 0$ but $x_0 \approx 3$, with slope $y_m = 1.61(5)$.}
\end{figure*}

The results we discussed in \secref{sec:phase} suggest that the transition at stronger coupling is related to chiral symmetry breaking, while the transition at weaker coupling is related to confinement.
We will consider confinement in the next section, while in this section we investigate the chiral properties of the phases.
Finite fermion mass explicitly breaks chiral symmetry, and extrapolating the chiral condensate to the $m=0$ chiral limit can be a difficult task.
Here we use the eigenvalue distribution of the Dirac operator to study chiral symmetry.

The spectrum of the Dirac operator of chirally broken systems contains a wealth of information.
When the eigenvalue distribution is compared to random matrix theory (RMT), it predicts the chiral condensate and gives information about the lattice artifacts of the simulations.
There is no comparable prediction for conformal systems, nevertheless the volume scaling and the level spacing of consecutive eigenvalues are related to the mass scaling exponent of the fixed point that governs the infrared dynamics~\cite{Fodor:2009wk, DeGrand:2009hu}.

The 12-flavor staggered action is local and describes a well-defined statistical system.
In the following analysis we will investigate general questions of scaling and the appropriate scaling dimension for this system.
We will not compare our data to RMT predictions, and our analysis will not be sensitive to taste symmetry, or its breaking.

In this pilot study we calculate the 12 lowest-lying eigenvalues of the staggered Dirac operator on volumes $12^4$, $12^3\X24$, $16^4$ and $16^3\X32$ in both the \Sb and weak-coupling phases.
In principle one should separate the different topological sectors before averaging the eigenvalues, but all of the configurations we analyzed appear to be in the zero-topology sector.
\fig{fig:eigenvalues} illustrates the volume dependence and the level spacing of the lowest-lying eigenvalues.
Our gauge configurations are too coarse for the eigenvalues to show the four-fold degeneracy expected in the continuum limit of staggered fermions.
Additional HYP smearing steps can remove enough of the ultraviolet fluctuations to reveal the expected degeneracy, but that is not what the dynamical fermions see in the simulations, and we do not pursue this direction.

In the infinite-volume limit the basic quantity is the eigenvalue density $\rho(\lambda)$.
In the chiral limit the density of low-lying eigenvalues is expected to scale as
\begin{equation}
  \label{eq:rho_lambda}
  \rho(\lambda) \propto (\lambda - \lambda_0)^\alpha,
\end{equation}
where the parameter $\lambda_0\ge 0$ allows the possibility of a soft edge~\cite{Bowick:1991ky, Akemann:1996vr, Damgaard:2000cx}.
In a chirally broken system $\rho(0) \ne 0$, implying $\lambda_0 = 0$ (the ``hard edge'') and $\alpha = 0$.
In a conformal system $\lambda_0 = 0$, and $\alpha$ is related to the mass anomalous dimension as we now derive.

Although the density $\rho(\la)$ is only well-defined in the infinite-volume limit, the functional form of \eqn{eq:rho_lambda} can be used to analyze the spacing between discrete eigenvalues in a finite volume.
Denoting the finite-volume eigenvalues as $\la_i$ for $i = 1$, 2, \dots, we can write the cumulative eigenvalue density as
\begin{equation}
  \label{eq:rho_lambda2}
  \int_{\tilde\la}^{\tilde\La} \rho(\lambda)d\lambda = \lim_{V \to \infty} \left(\frac{n - m}{V}\right),
\end{equation}
where $\lambda_n = \tilde\La$ and $\lambda_m = \tilde\la$.
Using \eqn{eq:rho_lambda} this leads to
\begin{equation}
  \label{eq:rho_lambda3}
  \lambda_n - \lambda_0 \propto \left(\frac{n - x_0}{V}\right)^{\frac{1}{\alpha+1}}\left[1 + {\mathcal O}\left(V^{-1}\right)\right],
\end{equation}
where we combined $m / V$ and $(\lambda_m - \lambda_0)^{\alpha + 1} \propto \mathcal O(V^{-1})$ in the parameter $x_0 / V$.
Since the eigenvalue $\lambda$ has dimension of mass, \eqn{eq:rho_lambda3} implies the relation
\begin{equation}
  \frac{D}{1 + \alpha} = y_m = 1 + \gamma_m
\end{equation}
between $\alpha$ and the mass anomalous dimension $\gamma_m$ in a $D$-dimensional space.
For free field theory $y_m = 1$ and $\alpha = D - 1$, while for a chirally broken system $\alpha = 0$.
\eqn{eq:rho_lambda3} has four free parameters.
The proportionality constant and $x_0$ can depend on the lattice geometry, while $\lambda_0$ and $\alpha$ are universal.

The left panel of \fig{fig:eigenvalues} shows our results for the low-lying eigenvalues in the \Sb phase at $\beta_F = 2.6$, $m = 0.005$.
In this plot we include a dashed line showing the soft edge $\la_0 = 0.0175(5)$ predicted by our global fit to \eqn{eq:rho_lambda3} using all four volumes.
The left panel of \fig{fig:eig_scaling} shows the result of this global fit, which does not depend on the aspect ratios of the lattices.
The dependence on $x_0$ is weak, and we fix $x_0 = 0$.
The slope $y_m = 4 / (\alpha + 1) = 2.50(10)$ gives $\alpha = 0.60(6)$, consistent with the RMT prediction $\alpha = 1 / 2$ at a soft edge~\cite{Bowick:1991ky, Akemann:1996vr}.

A non-vanishing soft edge is rather unusual.
In finite-temperature systems with $N_t$ fixed, $L \to \infty$, \refcite{Damgaard:2000cx} observed $\lambda_0 > 0$ in the chirally broken phase, but in infinite volume neither chirally broken nor conformal systems are expected to have a soft edge.
Through the Banks--Casher relation~\cite{Banks:1979yr}
\begin{equation}
  \label{eq:BanksCasher}
  \pbp \propto m \int_0^{\infty} \frac{\rho(\la)d\la}{\la^2 + m^2},
\end{equation}
a soft edge implies that the chiral condensate \pbp vanishes in the chiral limit $m = 0$.
With a soft edge, $\rho(\la) = 0$ for $0 \leq \la < \la_0$ as well as for all $\la$ larger than the spectral range of the Dirac operator, so that the integral in \eqn{eq:BanksCasher} remains finite while $m \to 0$.

In addition, a soft edge excludes the scenario in which $\pbp = 0$ but chiral symmetry is broken in the \Sb phase by a nonzero four-fermion condensate.
As discussed by Refs.~\cite{Stern:1997ri, Stern:1998dy, Kogan:1998zc}, this could result from the chiral symmetry breaking pattern
\begin{equation}
  \label{eq:pattern}
  SU(N_f)_V \X SU(N_f)_A \to SU(N_f)_V \X Z_{N_f}
\end{equation}
where the custodial $Z_{N_f}$ symmetry forces $\pbp = 0$.
The four-fermion condensate considered in \refcite{Kogan:1998zc} is related to the difference of scalar and pseudoscalar susceptibilities $\omega = \chi_P -\chi_S$ where
\begin{align}
  \chi_P & = \frac{1}{V} \int d^4x d^4y \vev{\chibar \tau^j i\ga_5 \chi(x) \chibar \tau^j i\ga_5 \chi(y)} \\
  \chi_S & = \frac{1}{V} \int d^4x d^4y \vev{\chibar \tau^j \chi(x) \chibar \tau^j  \chi(y)}
\end{align}
and $\tau^j$ is a flavor generator.
The U(1)$_A$-noninvariant $\omega$ parameter can be expressed in terms of the eigenvalue density as~\cite{Chandrasekharan:1998yx}
\begin{equation}
  \omega = 4m^2 \int_0^{\infty} \frac{\rho(\lambda)d\la}{(\lambda^2 + m^2)^2}\,.
\end{equation}
Just as for \eqn{eq:BanksCasher}, $\omega$ vanishes in the chiral limit if the eigenvalue density has a soft edge, so the symmetry breaking scenario of \eqn{eq:pattern} is not consistent with our data in the \Sb phase.

The soft edge is a dimensional parameter, but it is not clear what infinite-volume physical quantity it corresponds to.
Better understanding of the symmetry breaking mechanism could shed light on this question.

The eigenvalue spectrum in the weak-coupling phase is more conventional.
The right panel of \fig{fig:eigenvalues} shows the low-lying eigenvalues in this phase at $\beta_F = 2.7$, $m = 0.005$.
The global fit to \eqn{eq:rho_lambda3} predicts $\lambda_0 = 0$ but a non-vanishing $x_0 \approx 3$. 
Volumes with different aspect ratios prefer slightly different $x_0$ values and proportionality constants.
A fit with a common $\alpha$ parameter to the 12 eigenvalues on all four volumes predicts $y_m = 1.61(5)$ or mass anomalous dimension $\gamma_m = 0.61(5)$ (right panel of \fig{fig:eig_scaling}), where the error is purely statistical.
This value is consistent with results reported by Refs.~\cite{Appelquist:2011dp, DeGrand:2011cu, Aoki:2012kr}.
We find similar scaling properties and exponent at mass $m=0.01$.

In a chirally broken, confining system the eigenvalues should scale with exponent $y_m=4$, even in the $\epsilon$ regime where the volume is small compared to the pion Compton wavelength.
Our observation of uniform scaling on all four volumes indicates that the 12-flavor system is either conformal, or that its intrinsic confinement scale can only be observed on larger lattice volumes than we consider here.
In future work we will obtain a more robust prediction and quantify systematic effects by performing similar calculations at different gauge couplings and mass values, and on larger volumes as well~\cite{Cheng:2012TBD}.
For now, we turn to investigating the confinement properties of the \Sb and weak-coupling phases.

\section{\label{sec:pot_spec}Static potential and meson spectrum} 
\begin{figure}
  \centering
  \includegraphics[height=\PRDheight]{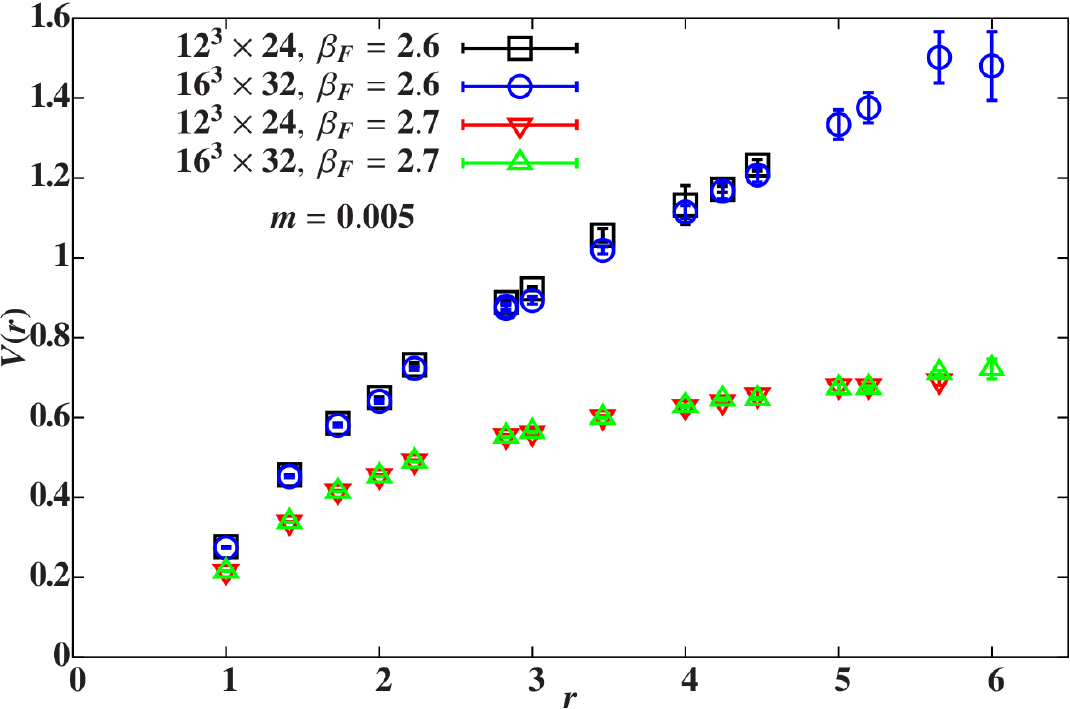}
  \caption{\label{fig:potential} The HYP-smeared static potential in the \Sb phase at $\be_F = 2.6$ and the weak-coupling phase at $\be_F = 2.7$.}
\end{figure}

In this section we explore the static potential and meson spectrum in the \Sb phase, contrasting these results with the same observables in the weak-coupling phase.
Although $\vev{\mbox{Tr}L_b}$ shows a clear signal in \fig{fig:pbp_Lb}, the Polyakov loop is not an order parameter in the presence of dynamical fermions.
The static potential is a more reliable indicator of confinement.
In \fig{fig:potential} we contrast the HYP-smeared static potential~\cite{Hasenfratz:2001tw} measured on each side of the transition, at $\be_F = 2.6$ and 2.7 on $12^3\X24$ and $16^3\X32$ volumes with $m = 0.005$.

The potential at $\be_F = 2.6$ is consistent with confinement, with string tension $\sigma=0.20(1)$ and Sommer parameter $r_0=2.67(4)$ in lattice units.
We obtain similar values at other masses and couplings within the \Sb phase.
The potential is almost identical on $12^3\X24$ and $16^3\X32$ volumes, and the small $r_0$ suggests that there will be no qualitative change on larger volumes that we are currently investigating.
These results indicate confinement with a fairly short gauge correlation length.
On the other hand, the potential at $\be_F = 2.7$ is coulombic and cannot be fitted consistently with a linear term.
The lack of volume dependence implies either vanishing string tension and conformality or an intrinsic confinement scale that can only be observed on larger lattice volumes.

\begin{figure*}[ht]
  \centering
  \includegraphics[width=0.45\linewidth]{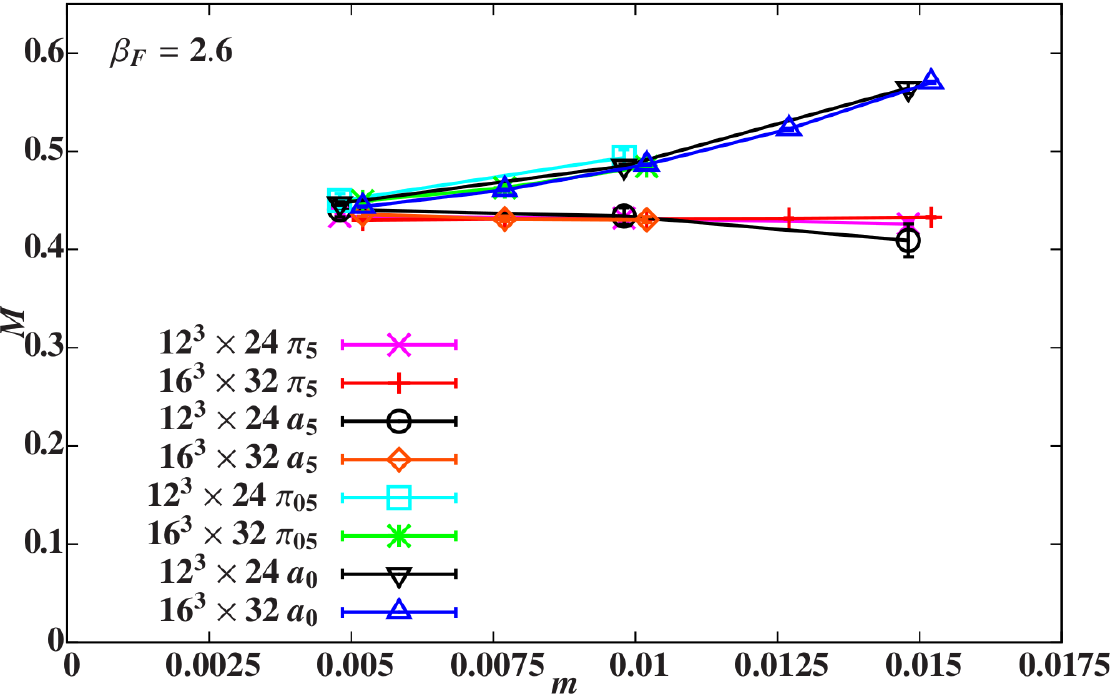}\hfill
  \includegraphics[width=0.45\linewidth]{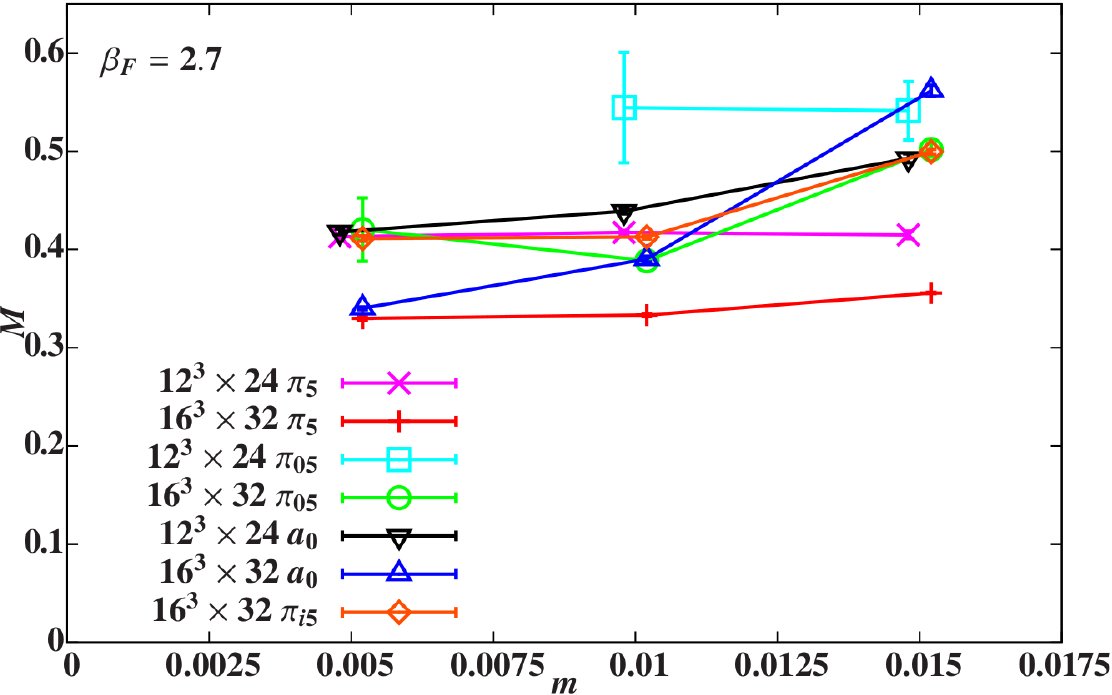}
  \caption{\label{fig:spectrum_vol} The masses of light scalar and pseudoscalar staggered mesons, from $12^3\X24$ and $16^3\X32$ lattices.    Small horizontal offsets distinguish results from different volumes.  In the left panel we include the Goldstone $\pi_5$, its ``$a_5$'' parity partner, the $\pi_{05}$ pseudoscalar and the $a_0$ scalar in the intermediate phase at $\be_F = 2.6$.  In the right panel we show the $\pi_5$, $\pi_{05}$, $a_0$ and (on $16^3\X32$ only) $\pi_{i5}$ in the weak-coupling phase at $\be_F = 2.7$.}
\end{figure*}

The meson spectrum at $\be_F = 2.7$ is also consistent with a small-volume deconfined system.
The right panel of \fig{fig:spectrum_vol} shows the Goldstone $\ga_5$ pseudoscalar ($\pi_5$), the pseudoscalar and the scalar components of the $\ga_0\ga_5$ channel ($\pi_{05}$ and $a_0$) and the $\ga_i\ga_5$ pseudoscalar ($\pi_{i5}$) versus fermion mass $m$.
We observe significant volume dependence in the scalar and pseudoscalar masses.
The Goldstone and the scalar become degenerate at small $m$, consistent with parity doubling.
The $\pi_{05}$ meson becomes heavier than the scalar at $m=0.005$, where it is degenerate with the $\pi_{i5}$ state.
(Our data do not allow precise results for these states on $12^3\X24$ at $m < 0.01$.)
Overall our meson spectrum results at $\be_F = 2.7$ are dominated by finite-volume effects, and do not provide clear information about the IR dynamics of the 12-flavor model.
With the computational resources available to us, we cannot compete with the large-volume spectral study of \refcite{Fodor:2011tu}.

Our goal in investigating the static potential and meson spectrum in the weak-coupling phase is to contrast these results with measurements in the \Sb phase, where we observe several interesting differences.
Our results for the pseudoscalar and scalar spectrum at $\be_F = 2.6$ are summarized in the left panel of \fig{fig:spectrum_vol}.
In the \Sb phase we find that the pion has a parity partner (``$a_5$'') in the $\ga_5$ channel, a state that is forbidden in QCD-like systems.
The masses measured on $16^3\X32$ and $12^3\X24$ volumes are indistinguishable in the \Sb phase: the finite volume corrections are negligible, consistent with the small correlation length indicated by the static potential.
The parity partner states both in the $\ga_5$ and $\ga_0 \ga_5$ channels are degenerate.
The $\ga_5$ states are largely independent of the fermion mass $m$ while the $\ga_0 \ga_5$ mesons' masses increase steadily with increasing $m$.
The data indicate that all four mesons could be degenerate in the chiral limit.
However, the $\pi_{05}$ mass again proved difficult to extract, and our statistics and volumes do not allow precise results for the $\pi_{05}$ at $m > 0.01$.

\begin{figure}
  \centering
  \includegraphics[height=\PRDheight]{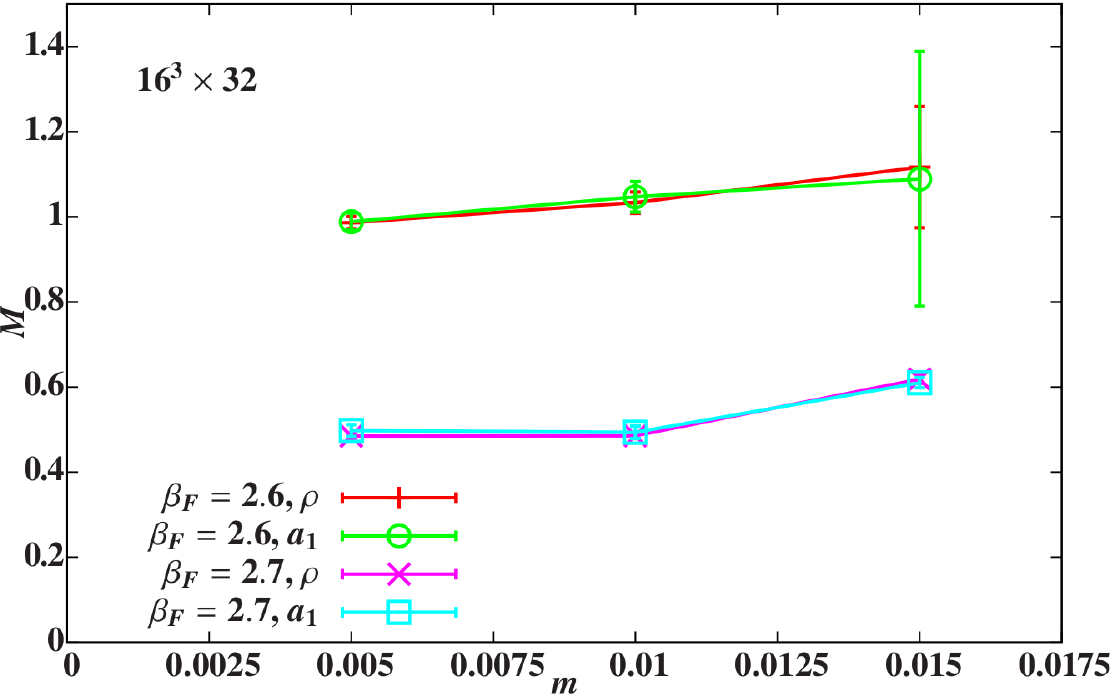}
  \caption{\label{fig:spectrum_VA} The $\rho$ and $a_1$ are degenerate in both the weak-coupling phase at $\be_F = 2.7$ as well as the \Sb phase at $\be_F = 2.6$.  At $\be_F = 2.7$ this degeneracy is a familiar effect.  At $\be_F = 2.6$, it is consistent with our observation of chiral symmetry in the eigenvalue spectrum.}
\end{figure}

In \fig{fig:spectrum_VA} we show the masses of the vector meson $\rho$ and its parity partner $a_1$ measured on $16^3\X32$ volumes in both the \Sb phase at $\be_F = 2.6$ and the weak-coupling phase at $\be_F = 2.7$.
At both of these couplings, the $\rho$ and $a_1$ are degenerate for all $m \leq 0.015$.
In the deconfined weak-coupling phase, this parity doubling is a familiar effect.
In the confining \Sb phase, however, such spectral properties are unusual.
The $\rho$--$a_1$ parity doubling we observe in \fig{fig:spectrum_VA} is inconsistent with the spectrum associated with the chiral symmetry breaking pattern of \eqn{eq:pattern}~\cite{Kogan:1998zc}.
Combined with the vanishing chiral condensate observed from the Dirac spectrum in this phase, the degeneracy of the parity partners in the meson spectrum implies that the intermediate phase is confining but chirally symmetric.
The continuum 't Hooft anomaly matching condition does not permit the existence of such a phase, suggesting that the novel phase we observe does not exist in the continuum.

\section{\label{sec:8f}The 8-flavor case} 
\begin{figure}
  \centering
  \includegraphics[height=\PRDheight]{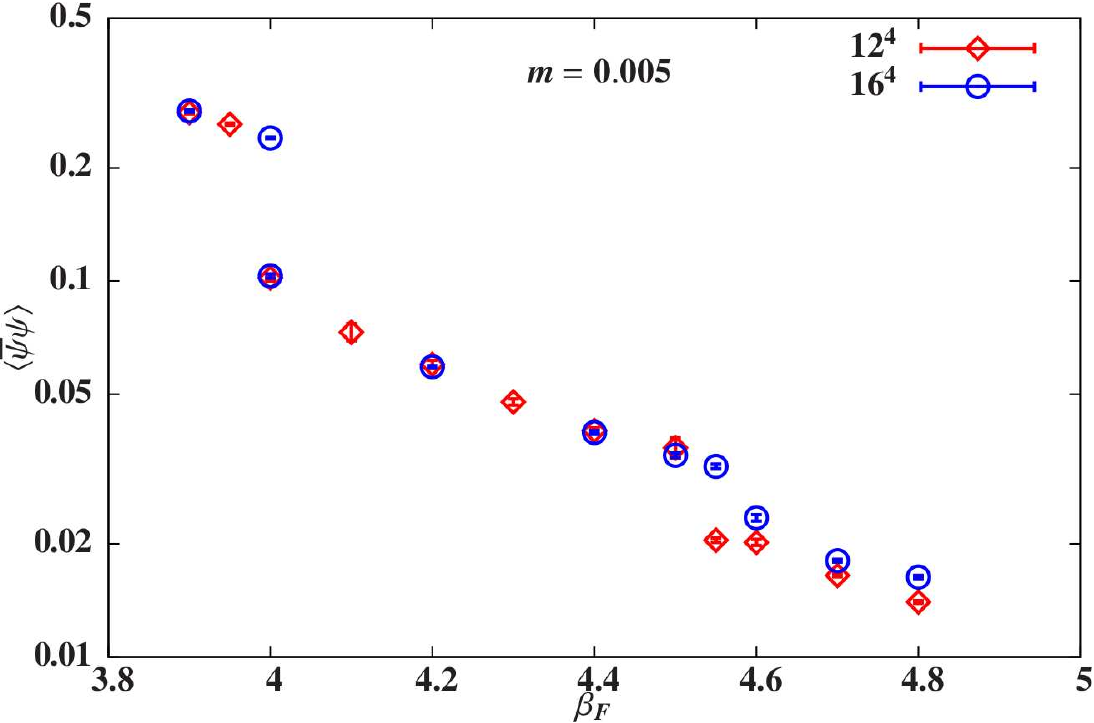}
  \caption{\label{fig:pbp_nf8} The chiral condensate \pbp (on a log scale) in the $N_f=8$ flavor system at $m=0.005$ on $12^4$ and $16^4$ lattices.  The phase between the two first order transitions is an \Sb phase like that we observe for $N_f = 12$.}
\end{figure}

Finite-temperature transitions converging to a bulk transition could signal that the continuum weak-coupling phase is conformal in the infrared.
However, because we observe two bulk transitions bounding an intermediate phase with unusual properties, we must consider the possibility that our results are due to lattice artifacts.
With Wilson fermions the existence of a lattice artifact phase, first proposed by Aoki~\cite{Aoki:1983qi}, is well known.
\refcite{Aubin:2004dm} argues that an Aoki-like phase might exist with staggered fermions if more than a single four-taste multiplet is considered.
We are currently investigating this possibility through additional studies with $N_f=4$, 8 and 16 flavors.

This work is preliminary, but important in interpreting our $N_f = 12$ results.
\fig{fig:pbp_nf8} shows the $N_f = 8$ chiral condensate \pbp at $m=0.005$ on $12^4$ and $16^4$ volumes.
We observe the same phases as with $N_f=12$ flavors.
On both volumes there are two first-order transitions at approximately volume-independent gauge couplings.
The phase in between has the same properties as the \Sb phase with 12 flavors.
It breaks single-site shift symmetry as shown by the non-zero expectation values of the two order parameters $\Delta P_{\mu}$ (\eqn{eq:plaq_diff}) and $\Delta L_{\mu}$ (\eqn{eq:link_diff}).
The Dirac operator eigenvalue spectrum has a soft edge, the static potential has a non-vanishing string tension, and the meson spectrum shows parity doubling.
Yet it is generally believed that the $N_f=8$ flavor system is below the conformal window~\cite{Deuzeman:2008sc, Appelquist:2009ty, Fodor:2009wk, Jin:2009mc, Jin:2010vm}, and our data in the weak-coupling phase support this expectation~\cite{Cheng:2012TBD}.

The fact that an \Sb phase exists with 8 flavors implies that this phase and its two corresponding bulk transitions do not necessarily imply IR conformality in the continuum.
The infrared behavior of the weak-coupling phase is independent of the \Sb phase and has to be studied by other means.

We emphasize that our treatment of the $N_f=8$ system is not complete, and our results with $N_f=4$ and 16 flavors are even more preliminary.
We include \fig{fig:pbp_nf8} to help clarify the situation with 12 flavors, but more work is needed to map out the full phase diagram.

\section{\label{sec:conclusions}Conclusion} 
Our investigations of the phase diagram of the 12-flavor SU(3) model have identified a novel phase with unusual properties.
At small masses this phase lies in between the usual confining, chirally broken lattice strong-coupling phase and the weak-coupling phase that is governed by the perturbative gaussian fixed point and possibly an infrared fixed point.
The two first-order phase transitions separating these three phases get closer together with increasing fermion mass.
At some mass value the two transitions merge and eventually turn into a crossover.
The intermediate phase forms a packet in between the strong- and weak-coupling phases at small fermion masses.

In this work we studied the intermediate phase and contrasted it with the weak-coupling phase using several observables.
The chiral condensate \pbp and blocked Polyakov loop gave our first glimpse of the phase structure, and suggested that the transition at stronger coupling is related to chiral symmetry breaking, while the transition at weaker coupling is related to confinement.
Our investigation led us to two operators, $\Delta P_{\mu}$ (\eqn{eq:plaq_diff}) and $\Delta L_{\mu}$ (\eqn{eq:link_diff}), that serve as order parameters of the intermediate phase.
Both of these order parameters are sensitive to the single-site shift symmetry ($S^4$) of the staggered fermions, a symmetry that is exact at the level of the lattice action.
Since these order parameters develop non-zero expectation values in the intermediate phase, but vanish in both the strong- and weak-coupling phases, we conclude that the intermediate phase spontaneously breaks single-site shift symmetry, $\Sb$.
Since the single-site shift symmetry is exact even at finite fermion mass, the \Sb phase must be separated by real phase transitions from both the strong- and weak-coupling phases.

We used the spectrum of the Dirac operator to study the chiral properties of the phases.
In the \Sb phase we found evidence for a soft edge, implying chiral symmetry.
In the weak-coupling phase the eigenvalue spectrum is consistent with both conformal and volume-squeezed confining scenarios.
We obtained a preliminary prediction for the mass anomalous dimension, $\gamma_m=0.61(5)$ in the weak-coupling phase.

The static potential showed that the \Sb phase is confining with a small lattice correlation length, while in the weak-coupling phase on our relatively small volumes the potential was only coulombic.
These results are consistent with the signal from the (blocked) Polyakov loop.
The meson spectrum in both phases indicated parity doubling at light fermion mass.
However, in the \Sb phase we observed very little volume dependence, yet we found that all mesons remained massive in the chiral limit.
The parity doubling in the weak coupling phase was accompanied by strong volume dependence and could also be consistent with both conformal and confining scenarios.

We presented preliminary data showing that the \Sb phase is present with 8 flavors as well, suggesting that this phase is not related to conformal infrared dynamics.
Our findings lead us to believe that the \Sb phase is a lattice artifact of the staggered fermions~\cite{Aubin:2004dm}.
Since the single-site shift symmetry is closely related to the fermion staggering and taste breaking, it is most likely that the origin of the \Sb phase is in the fermionic sector.
However, we do not yet have a clear and complete understanding of the symmetry breaking mechanism that produces the \Sb phase.

Further investigations of the phases on larger volumes, and with $N_f=4$, 8 and 16 fermions, are under way and should clear up the still open questions of these surprisingly complex systems.

\section*{Acknowledgments} 
We thank D.~Kaplan, D.~Son, T.~DeGrand, S.~Sharpe, J.~Kuti and P.~Damgaard for helpful comments and discussions.
A.~H.\ is indebted to S.~Schaefer for suggesting to change the smearing parameters to control the numerical problems of nHYP fermions.
Part of this work was completed while A.~H.\ was at the Kavli Institute for Theoretical Physics, supported in part by the U.S.~National Science Foundation (NSF) under Grant No.~NSF PHY11-25915.
A.~H.\ and D.~S.\ also benefited greatly from the 2012 Kobayashi--Maskawa Institute mini-workshop on ``Conformality in Strong Coupling Gauge Theories at LHC and Lattice'' in Nagoya, Japan.
We are grateful in particular to J.~Kuti for drawing our attention to the possibility of an Aoki-like phase for staggered fermions.

This research was partially supported by the U.S.~Department of Energy (DOE) through Grant No.~DE-FG02-04ER41290.
Numerical calculations were carried out on the HEP-TH and Janus clusters at the University of Colorado; at Fermilab under the auspices of USQCD supported by the DOE SciDAC program; and at the Texas Advanced Computing Center through the Extreme Science and Engineering Discovery Environment supported by NSF Grant No.~OCI-1053575.

\bibliographystyle{apsrev}
\bibliography{12f_phase_letter}
\end{document}